\documentclass[]{ceurart}

\DeclareOldFontCommand{\tt}{\normalfont\ttfamily}{\mathtt}

\begin{document}

\copyrightyear{2024}
\copyrightclause{Copyright for this paper by its authors.\\
  Use permitted under Creative Commons License Attribution 4.0
  International (CC BY 4.0).}

\conference{CHR 2024: Computational Humanities Research Conference, December 4-6, 2024, Aarhus, Denmark}

\title{Integrating Visual and Textual Inputs for Searching Large-Scale Map Collections with CLIP}

\author[1]{Jamie Mahowald}[%
orcid=0009-0007-1731-497X,
]
\ead{j.mahowald@utexas.edu}
\ead[url]{https://github.com/j-mahowald}
\address[1]{The Archer Center \& Department of Mathematics, The University of Texas at Austin, USA}

\author[2]{Benjamin Charles Germain Lee}[%
orcid=0000-0002-1677-6386,
]
\address[2]{Information School, University of Washington, USA}
\ead{bcgl@uw.edu}
\ead[url]{https://www.bcglee.com/}


\begin{abstract}
Despite the prevalence and historical importance of maps in digital collections, current methods of navigating and exploring map collections are largely restricted to catalog records and structured metadata. In this paper, we explore the potential for interactively searching large-scale map collections using natural language inputs (``maps with sea monsters''), visual inputs (i.e., reverse image search), and multimodal inputs (an example map + ``more grayscale''). As a case study, we adopt 562,842 images of maps publicly accessible via the Library of Congress's API. To accomplish this, we use the mulitmodal Contrastive Language-Image Pre-training (CLIP) machine learning model to generate embeddings for these maps, and we develop code to  implement exploratory search capabilities with these input strategies. We present results for example searches created in consultation with staff in the Library of Congress's Geography and Map Division and describe the strengths, weaknesses, and possibilities for these search queries. Moreover, we introduce a fine-tuning dataset of 10,504 map-caption pairs, along with an architecture for fine-tuning a CLIP model on this dataset. To facilitate re-use, we provide all of our code in documented, interactive Jupyter notebooks and place all code into the public domain. Lastly, we discuss the opportunities and challenges for applying these approaches across both digitized and born-digital collections held by galleries, libraries, archives, and museums.
\end{abstract}

\begin{keywords}
  maps \sep
  Library of Congress \sep
  computing cultural heritage \sep
  multimodal machine learning \sep \\
  exploratory search
\end{keywords}

\maketitle

\section{Introduction}

\begin{figure}
\begin{center}
    \includegraphics[width = \linewidth]{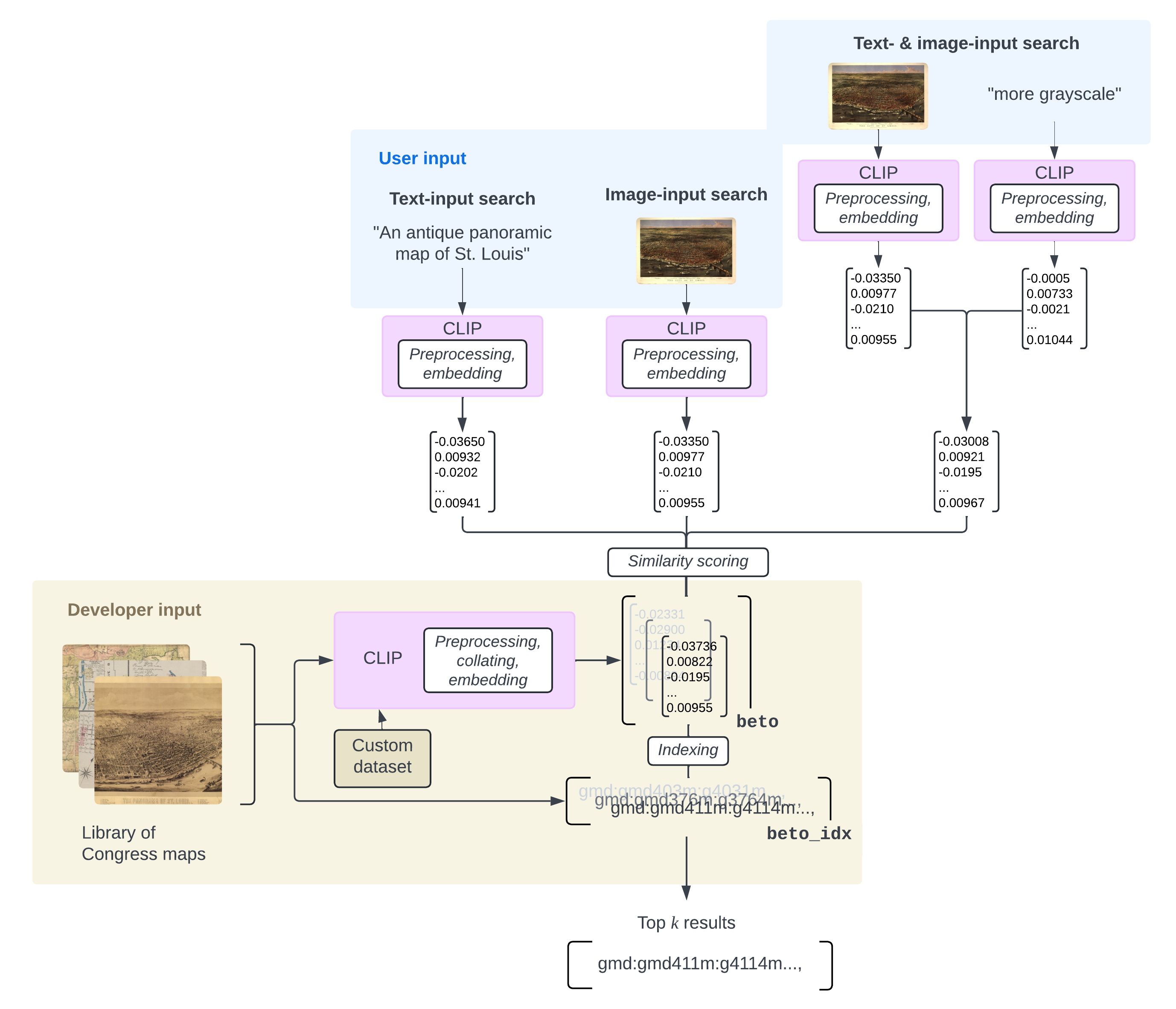}
\end{center}
\caption{An overview of our search implementation for 562,842 images of maps held by the Library of Congress. Users can create three different types of inputs: 1) text, 2) image, and 3) text \& image. The specified search input is used to retrieve the most relevant maps by dynamically computing a CLIP embedding for the search input and comparing it to the pre-computed CLIP embeddings of all 562,842 images (more information on the \texttt{beto} architecture used for this step can be found in Section \ref{sec:methodology}). The top $k$ search results are then returned to the user.}
\label{fig:overview}
\end{figure}

Maps represent a central collecting focus for cultural heritage institutions, comprising large fractions of collections across the world. For example, the Library of Congress alone holds over 5.5 million maps \cite{zhang_2012}. Efforts to digitize maps have resulted in new possibilities for access for a wide range of patrons, from scholars to politicians to the public. However, current methods for searching historic map collections are largely limited to structured metadata and keyword search over extracted text via optical character recognition (OCR). As described in a 2020 survey of metadata for topographic maps, metadata is ``not often connected with the way in which users search for maps,'' and metadata standards vary across institution 
\cite{topoographic_metadata}. Furthermore, enriching metadata requires staff time and expertise, which is not always feasible.

In recent years, scholars and practitioners within cultural heritage, the computational humanities, and the digital humanities have begun exploring the application of computer vision methodologies to historic maps for a wide range of tasks, ranging from metadata enrichment via classification \cite{schnurer} to the semantic identification of visual markers such as railroad tracks \cite{mapreader}. However, much remains to be explored surrounding methods for facilitating exploration and sensemaking of map corpora using machine learning.

In this paper, we take on this challenge by exploring possibilities for searching large-scale map collections using multimodal machine learning. As a case study, we adopt as our collection of choice 562,842 images of maps publicly available through the Library of Congress's API. To facilitate multimodal search and discovery, we generate and release embeddings for these images using OpenAI's CLIP model \cite{CLIP_paper}. Significantly, CLIP and other multimodal approaches have seen increasing adoption in the computational humanities community, showing great promise for use with digital collections \cite{wevers_photos}. We build on this work to show the possibilities for maps in particular. Namely, we leverage the shared image- and text-embedding space enabled by CLIP to implement three different forms of interactive search: with natural language inputs (``maps with sea monsters''), with visual inputs (i.e., reverse map search), and with multimodal inputs (an example map + ``more grayscale''). Our search code is highly responsive, capable of searching half a million images and returning results on a consumer-grade GPU (e.g., using a personal laptop) in less than a second. We also introduce a dataset of 10,504 map-caption pairs, as well as code for fine-tuning CLIP with this dataset. To facilitate re-use, all of our code is released into the public domain in the form of documented Jupyter notebooks that others can run on their own machines. These notebooks can be found at our GitHub repository: \url{https://github.com/j-mahowald/clip-loc-maps}. An overview of our search implementation can be found in Figure \ref{fig:overview}. 

Working in collaboration with the Library of Congress's Geography and Map division, we present a number of example searches using our search implementation and describe the strengths and limitations of this approach. Given the shared challenges surrounding discoverability  across digital collections, we discuss the extensibility of these results to other cultural heritage collections, ranging from digitized materials to born-digital content. In order to ensure that our work has been conducted ethically and responsibly, we describe our adoption of the LC Labs AI Planning Framework throughout our research process.

In summary, our paper offers five central contributions:
\begin{enumerate}
\item We introduce CLIP embeddings for 562,842 images of 56,554 map items held by the Library of Congress and made available through the loc.gov API.
\item We introduce a dataset of 10,504 map-caption pairs, as well as an architecture for fine-tuning a CLIP model on this dataset.
\item In consultation with the Geography and Map Division at the Library of Congress, we demonstrate the utility of these embeddings for a range of search \& discovery tasks, including natural language search, reverse image search, and multimodal search.
\item We release all of our code as re-usable Jupyter notebooks and place the notebooks into the public domain. These notebooks include our pipeline for generating the CLIP embeddings, our search implementation for all three methods, and the code for fine-tuning CLIP. Our code can be found in our \href{https://github.com/j-mahowald/clip-loc-maps}{GitHub repository}.
\item We discuss potential ways that CLIP embeddings could be used to improve discoverability across digital collections.
\end{enumerate}
\section{Related Work}

In recent years, the ``Collections as Data'' movement and related efforts have demonstrated the value of applying artificial intelligence (AI) to digital collections held by galleries, libraries, archives, and museums in a range of contexts \cite{collections_as_data_2017, collections_as_data_2019}. Of particular relevance to this paper is work that has applied computer vision to digital collections in the context of search and discovery \cite{pixplot, kblab, nationalneighbors, tsnemap, principalcomponents, vane, amnh, anadol, visualturn, nn_dataset, nn_demo}. Likewise, the MapReader project and others have demonstrated value in applying machine learning to historic maps for classification and other tasks \cite{schnurer, mapreader, vitale2023Searching, uhl_2018, uhl_2022}. In this paper, we pursue the intersection of these bodies of work in order to explore the application of machine learning methods to search and discovery for historic maps. Surrounding the use of multimodal machine learning approaches, we build on work by Smits \& Wevers \cite{multimodal_turn}, Smits \& Kestemont \cite{smits_kestemont}, and Barancová et al. \cite{wevers_photos}. This collective work has applied OpenAI's CLIP model \cite{CLIP_paper} to digital collections, and we follow suit, focusing on the application to map collections in particular. Though work such as PIGEON \cite{haas2023pigeon} and StreetCLIP \cite{streetCLIP} have applied multimodal machine learning approaches to maps, the focus has been on contemporary, born-digital maps, whereas we consider historic maps and give attention to the specific context of cultural heritage.

Significantly, much work has explored the availability and usability of metadata for historic map collections \cite{topoographic_metadata, Janes_2012, maps_lit_review, maprank}. In this paper, we ask how search and discovery can be enriched beyond existing metadata; we refer to this literature for further reading on the strengths and limitations of existing metadata practices.

Lastly, our work is situated within a landscape of research actively engaging with the responsible and ethical dimensions of applying AI to cultural heritage collections. We note that many frameworks and guidelines exist for pursuing this work \cite{padilla_responsible_2020, cordell_machine_nodate, collections_as_ml_data, eventsummaryLCLabs, johd}. In this paper, we adopt the LC Labs AI Planning Framework in particular because we utilize Library of Congress maps for our case study \cite{planningframework}. In Section \ref{sec:dataset}, we describe our dataset in more detail.

\section{Methodology 1: Generating CLIP Embeddings for 562,842 Images of Maps}

\subsection{Our Dataset of Library of Congress Maps}\label{sec:dataset}

The Library of Congress has made publicly available over 56,000 map items comprising over 563,000 segments (images), a figure that continues to grow regularly. The map items are largely from the Geography and Map (G\&M) Division and are vastly varied in relation to the number of constituent images included, with some containing only one image (e.g., maps of small towns, or standalone illustrations), while others, such as atlases or set maps, contain orders of magnitude more. For example, the Texas General Highway Map item contains over 10,000 sheets. Of the 563,696 segments we attempted to process, 562,842 (99.85\%) returned valid requests through the International Image Interoperability Framework (IIIF) when we queried them for our purposes. Each item is associated with one or more resources, onto which individual segments add an identifying suffix. For instance, the resource \texttt{g4031pm.gct00608}, which represents the first 2,999 sheets of a map set named ``General highway map ... Texas,'' includes \texttt{g4031pm.gct00608.cs000150} representing a particular sheet showing highways in Aransas County.

\subsection{Generating Embeddings}\label{sec:methodology}

\begin{figure}
\begin{center}
 \includegraphics[width=0.85\linewidth]{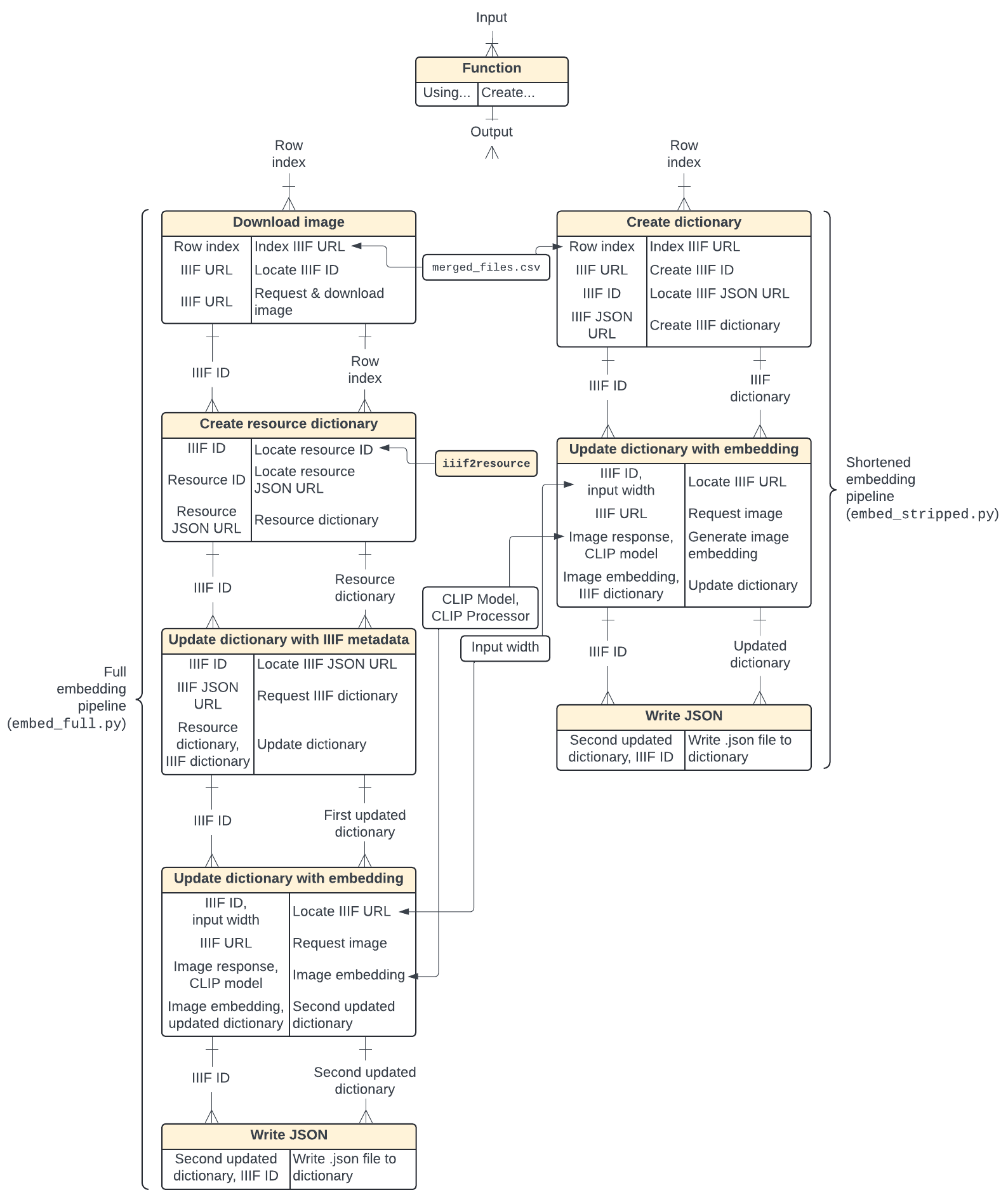}   
\end{center}
\caption{An overview of our embeddings generation pipeline along with the resulting metadata dictionaries, as described in this section. Each box represents a function; each function's left column lists objects created, while the right column lists the method of creation. Arrows between boxes represent items piped through successive functions.}
\label{fig:pipeline}
\end{figure}

We introduce a pipeline that leverages multiprocessing to efficiently generate embeddings for the Library of Congress maps in our dataset, while retaining their metadata and structure (Figure \ref{fig:pipeline}). Our embeddings generation pipeline can process over half a million images on an M3 MacBook Pro with 18GB memory in under 24 hours. 

To generate the embeddings using the base CLIP model, we tested a range of image widths and patch sizes, settling on width \texttt{w=2000px} and the base-size Vision Transformer (ViT) with \texttt{32x32px} patches to optimize download times while retaining sufficiently high image resolution. We then built out our pipeline around these specifications.

Of the several forms of identification that belong to each Library of Congress object, we focus on the IIIF ID for image processing and unique identification and the resource ID for metadata extraction. The data frame \texttt{merged\_files.csv} –– accessible via \href{https://zenodo.org/records/11538437}{Zenodo} \cite{mahowald_2024_11538437} –– gives the \texttt{loc.gov} resource URL and IIIF image URL for each object. It also provides information on an image's file size and its context in the collection. The pipeline reads row by row from this CSV file, creating a metadata dictionary for each image that includes API metadata. Using the defined preprocessor, model, and IIIF request, embeddings are generated, normalized, and appended to this dictionary as an \(m=512\)-tuple. 

Our GitHub repository contains two script versions: \texttt{embed\_full.py}, which incorporates extensive metadata from the \texttt{loc.gov} API, ideal for further fine-tuning, and \texttt{embed\_stripped.py}, which includes only the IIIF image URL and its embedding. Each IIIF ID can be used to derive its corresponding resource ID (the converse is not true), so only the IIIF ID is strictly needed to carry the pipeline forward. The JSON files are then written to a local directory and named for their IIIF ID to ensure uniqueness and easy derivation.

\begin{figure}
\begin{center}
\includegraphics[width = 0.9\linewidth]{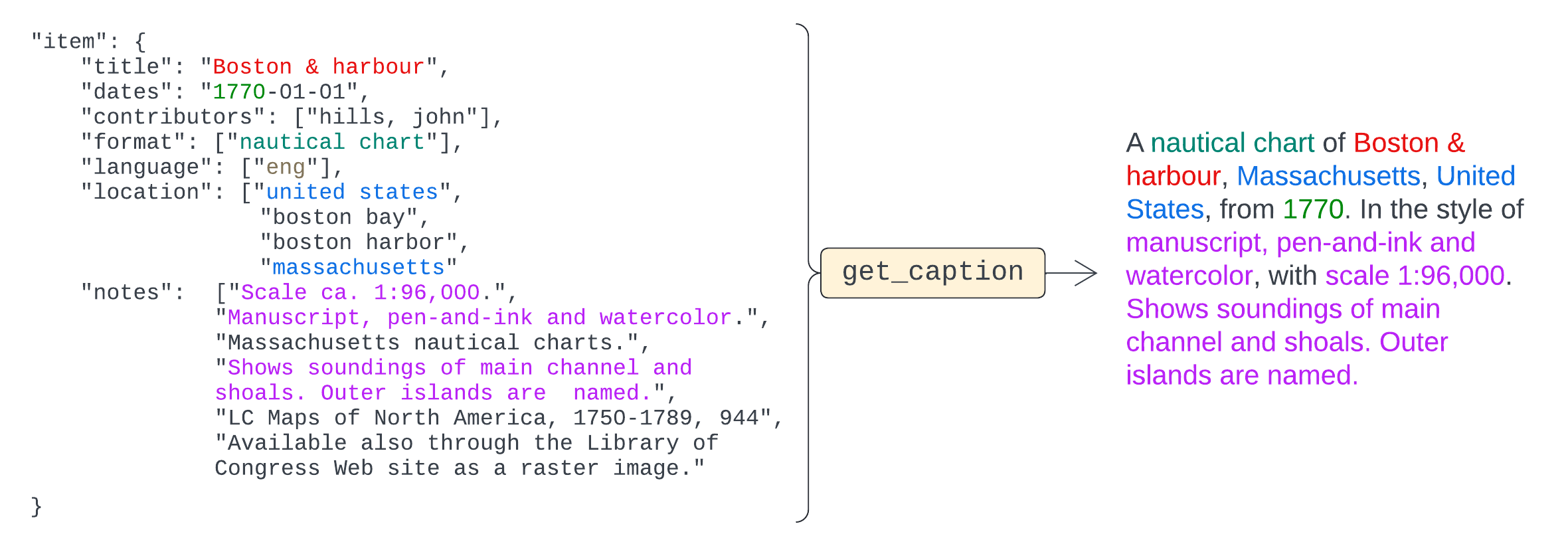}
\end{center}
\caption{A natural language caption is extracted from the \texttt{loc.gov} metadata through the \texttt{get\_caption} function without the need for a language model. The function is designed to avoid redundancy –– in this instance, it skips over the entries in \texttt{"location"} that include terms already extracted from the title. Metadata notes tend to be formatted such that only the first several items contain visual feature information, while the last two notes contain catalogue and availability data less pertinent to feature extraction.} 
\label{fig:natural-language-caption}
\end{figure}

After each JSON is downloaded, \texttt{create\_beto.py} generates \texttt{beto} (``\textbf{b}ig \textbf{e}mbedding \textbf{t}ensor \textbf{o}bject''), a PyTorch tensor of size \( [(m,), n] \) for \(n\) image embeddings of dimension \(m\). Though only two-dimensional, the tensor formulation facilitates indexing and serves as an input for a search query. A corresponding \(n\)-tuple, \texttt{beto\_idx}, is created to associate each embedding in \texttt{beto} with its respective IIIF URL by index. Diagrammatically, 

\begin{equation}\label{eq:beto}
\texttt{beto} =
\begin{bmatrix}
    \begin{bmatrix}
        a_{11} \\ a_{21} \\ ... \\ a_{m1}
    \end{bmatrix},
    & 
    \begin{bmatrix}
        a_{12} \\ a_{22} \\ ... \\ a_{m2}
    \end{bmatrix},
    & 
    ...,
    &
    \begin{bmatrix}
        a_{1n} \\ a_{2n} \\ ... \\ a_{mn}
    \end{bmatrix}
\end{bmatrix},
\qquad 
\texttt{beto\_idx}=
\begin{bmatrix}
    L_1, & L_2, & ..., & L_n
\end{bmatrix}
\end{equation}
where the column vector \([a_{1i}, a_{2i}, ..., a_{mi}]\) represents the \(m\)-tuple embedding for the \(i\)th image, and \(L_i\) represents the \(i\)th image's corresponding IIIF URL.

\subsection{A Dataset \& Architecture for Fine-tuning}\label{sec:finetuningdataset}

To complement the CLIP embeddings that we have generated and released, we introduce a dataset of 10,504 map-caption pairs for fine-tuning CLIP (available in  \href{https://zenodo.org/records/11538437}{Zenodo} \cite{mahowald_2024_11538437}), along with code for performing this fine-tuning (available in our \href{https://github.com/j-mahowald/clip-loc-maps/tree/main/fine-tuning}{GitHub repository}).

\begin{figure}
\begin{center}
    \includegraphics[width=\linewidth]{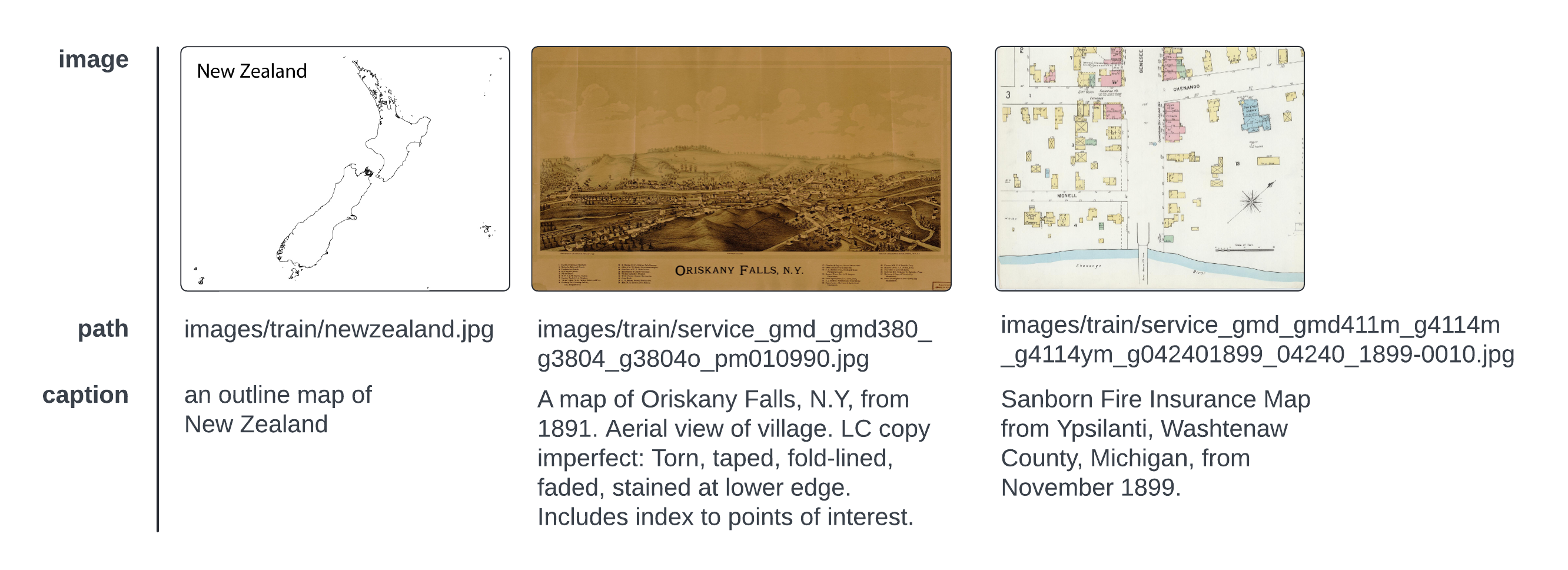}
\end{center}
\caption{Three examples of the map-caption pairs from our resulting dataset.}
\label{fig:three-pairs}
\end{figure}

One central goal in fine-tuning is to provide the CLIP model with a large set of map-caption (i.e., image-text) pairs from which it can contrastively learn relevant information such as styles, locations, dates, and other visual features. For each resource ID, the Library of Congress catalog record yields several descriptors useful for systematically training a model on map-caption pairs. We smooth the capitalization and punctuation through a few simple functions, and we use this metadata to generate a descriptive, natural language caption for each map. An example of the process is shown in Figure \ref{fig:natural-language-caption}, and three resulting examples are shown in Figure \ref{fig:three-pairs}. In total, we include 10,504 map-caption pairs by initially generating 10,000 maps with a single associated image, adding 2,000 randomly sampled images from the Sanborn Maps collection (which represents a disproportionate fraction of map images made publicly available online by the Library of Congress), adding an additional 227 maps covering every present-day country and U.S. state, and discarding a total of 1,723 samples with unresponsive image requests or low-quality captions (for instance, those with nonsensical characters, arbitrary changes in language, or no feature descriptors).

Our initial experiments performing our fine-tuning yielded mixed results. In Section \ref{sec:finetuning} in the Appendix, we describe these experiments. We also elaborate on our choice of our fine-tuning dataset.
\section{Methodology 2: Implementing Search \& Discovery with CLIP Embeddings}\label{sec:methodology_2}

In this section, we outline our implementation of our three different search methods using CLIP embeddings: 1) text-input search, 2) image-input search, and 3) text- \& image-input search. 

\subsection{Text-input Search}
Given a text query and a specified number of desired results \(k\), we employ the following process to search \texttt{beto} and \texttt{beto\_idx}, as defined in Equation \ref{eq:beto}. First, we generate a normalized text embedding for the query utilizing the same CLIP configuration used in encoding the larger collection. This embedding is then used to compute cosine similarity scores with each embedding in \texttt{beto}. We then identify the \(k\) largest scores and their corresponding indices. The similarity scores of these top \(k\) results are normalized using the softmax function, and the resulting scores, along with their respective identifying links from \texttt{beto\_idx}, are displayed. Cosine similarity is extremely efficient to compute, making it possible to identify the top \(k\) scores among over half a million images nearly instantaneously.

\subsection{Image-input Search}
In this strategy, the user can input an image URL and a desired number of results \(k\) to conduct this image-input search, alternatively known as reverse image search. After the URL request is received, the process is identical to the one outlined in text-input search because CLIP embeds images and text in a common embedding space. The query image is embedded on the spot as part of the search script, meaning that the user can input any image of any size and is not limited to those from the Library of Congress catalog. 

\subsection{Text- \& Image-input Search}

We introduce an experimental search strategy that accepts both a text string and image input as a search query. The CLIP model embeds the text string and the image to the same \(m\)-dimensional embedding space. The engine then accepts a scaling factor $\alpha$ that determines how much relative weight should be assigned to the text and image inputs. We assign: 
\begin{equation}
\textbf{c} = \frac{(1 - \alpha) \cdot \textbf{q} + (1 + \alpha) \cdot \textbf{t}}{2},\qquad \alpha \in [-1, 1]
\end{equation}
where \(\textbf{q}\) and \(\textbf{t}\) are the \(m\)-dimensional embedding for the image and text queries, respectively, and \(\textbf{c}\) is the combined weighted embedding (intuitively, this is a weighted centroid in the embedding space whose weights are determined by the scaling factor). Introducing this scaling factor satisfies the desired qualities that:
\begin{enumerate}
    \item an input of \(\alpha = 0\) weighs each term equally,
    \item the weight produced by a scaling term is equal to the reciprocal of the weight produced by the negative of that scaling term (i.e., a positive input \(\alpha_0\) weighs in favor of the text input exactly as much as \(-\alpha_0\) weighs in favor of the image input), and
    \item \(\textbf{c}\) limits to the sole input of \(\textbf{q}\) or \(\textbf{t}\) as \(\alpha\) approaches -1 or 1, respectively.
\end{enumerate} 
Our search engine then computes cosine similarity scores between this combined embedding and each embedding in \texttt{beto}, returning the top \(k\) scores as input by the user.

\section{Results \& Discussion}

In this section, we introduce example search results for all three strategies described in Section \ref{sec:methodology_2} and reflect on the strengths and limitations of our implementation. We also describe our utilization of the LC Labs AI Planning Framework throughout our research process.

\begin{figure}[pos=htbp]
\begin{center}
\textbf{Text-input Search}\\
 \includegraphics[width = 0.74\linewidth ]{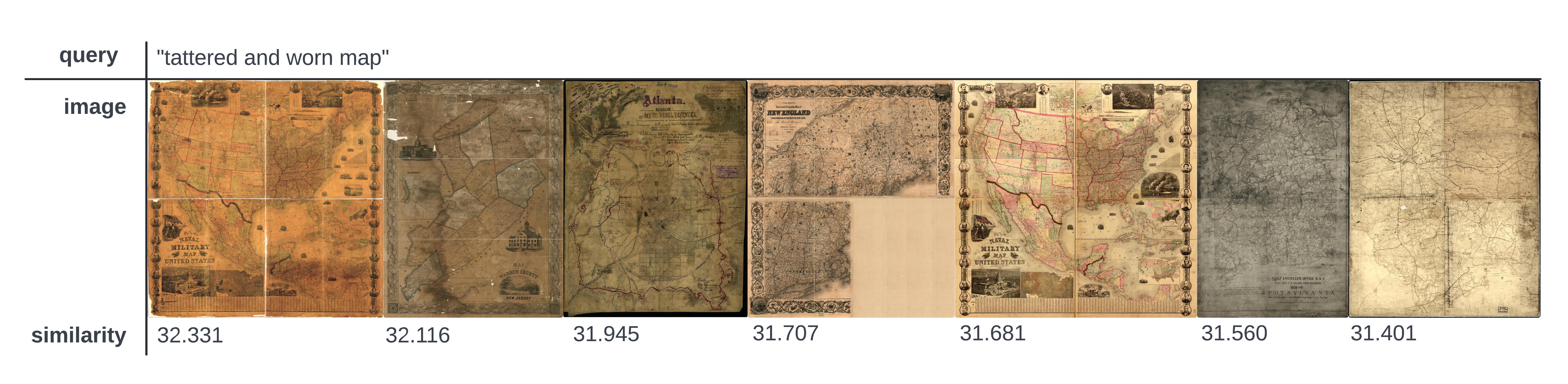} 
 \includegraphics[width = 0.74\linewidth ]{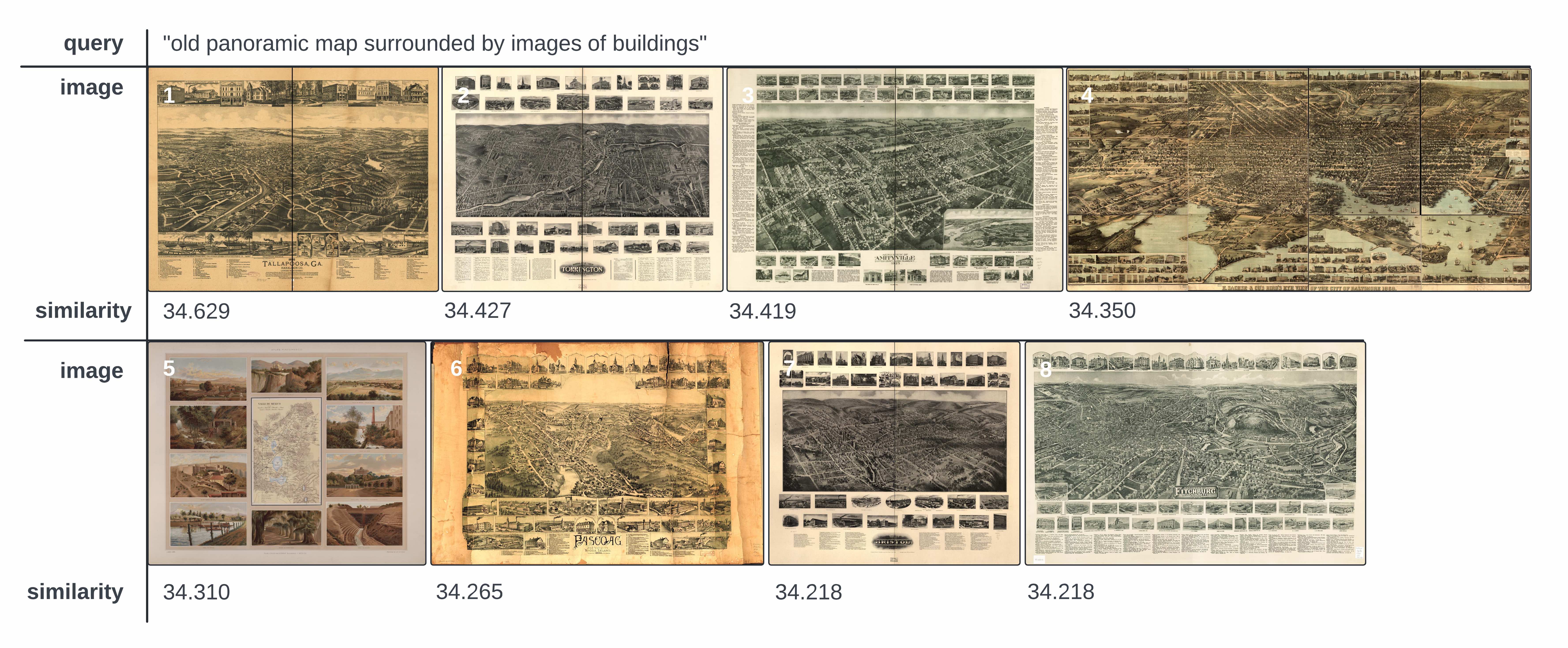} 
 \includegraphics[width = 0.74\linewidth ]{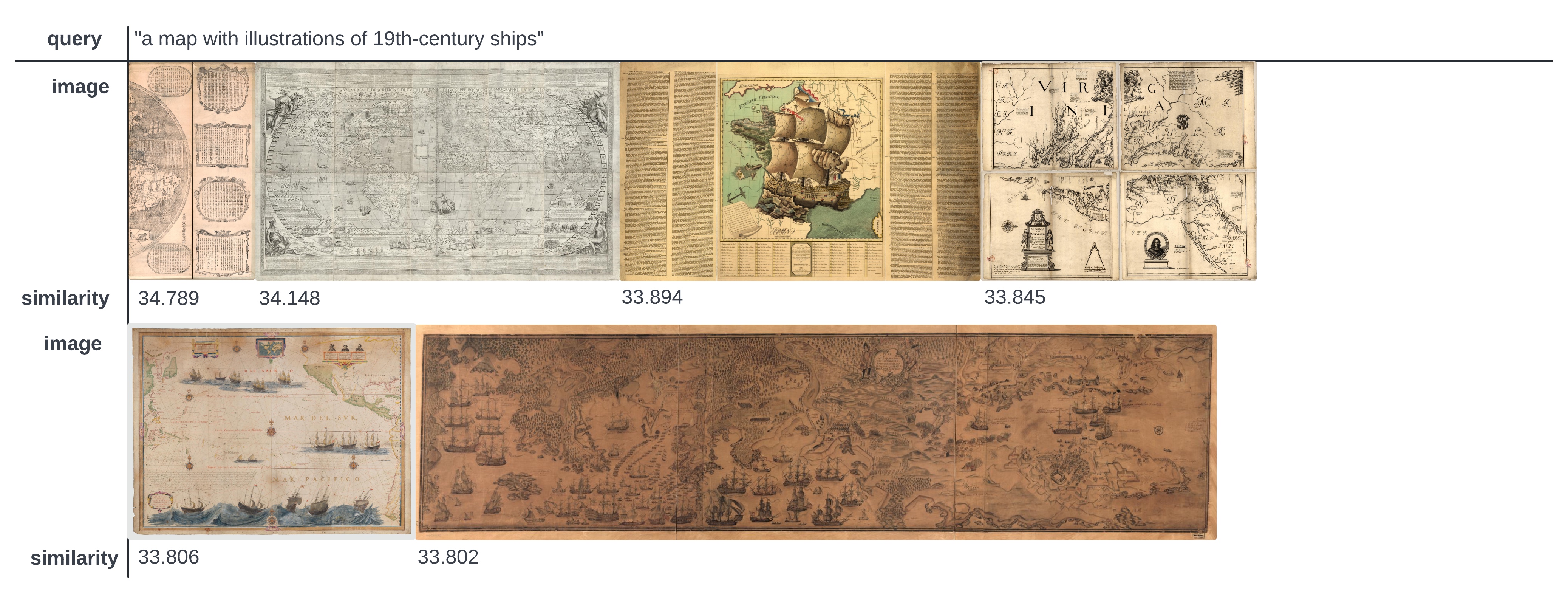}
\end{center}
\caption{The most relevant images returned by our search system for the following text queries: ``tattered and worn map,'' ``old panoramic map surrounded by images of buildings,'' and ``a map with illustrations of 19th-century ships.'' Non-normalized similarity scores are shown for each returned image.
}
\label{fig:results_text}
\end{figure}

\begin{figure}[pos=htbp]
\begin{center}
\textbf{Image-input Search}\\
 \includegraphics[width = 0.74\linewidth ]{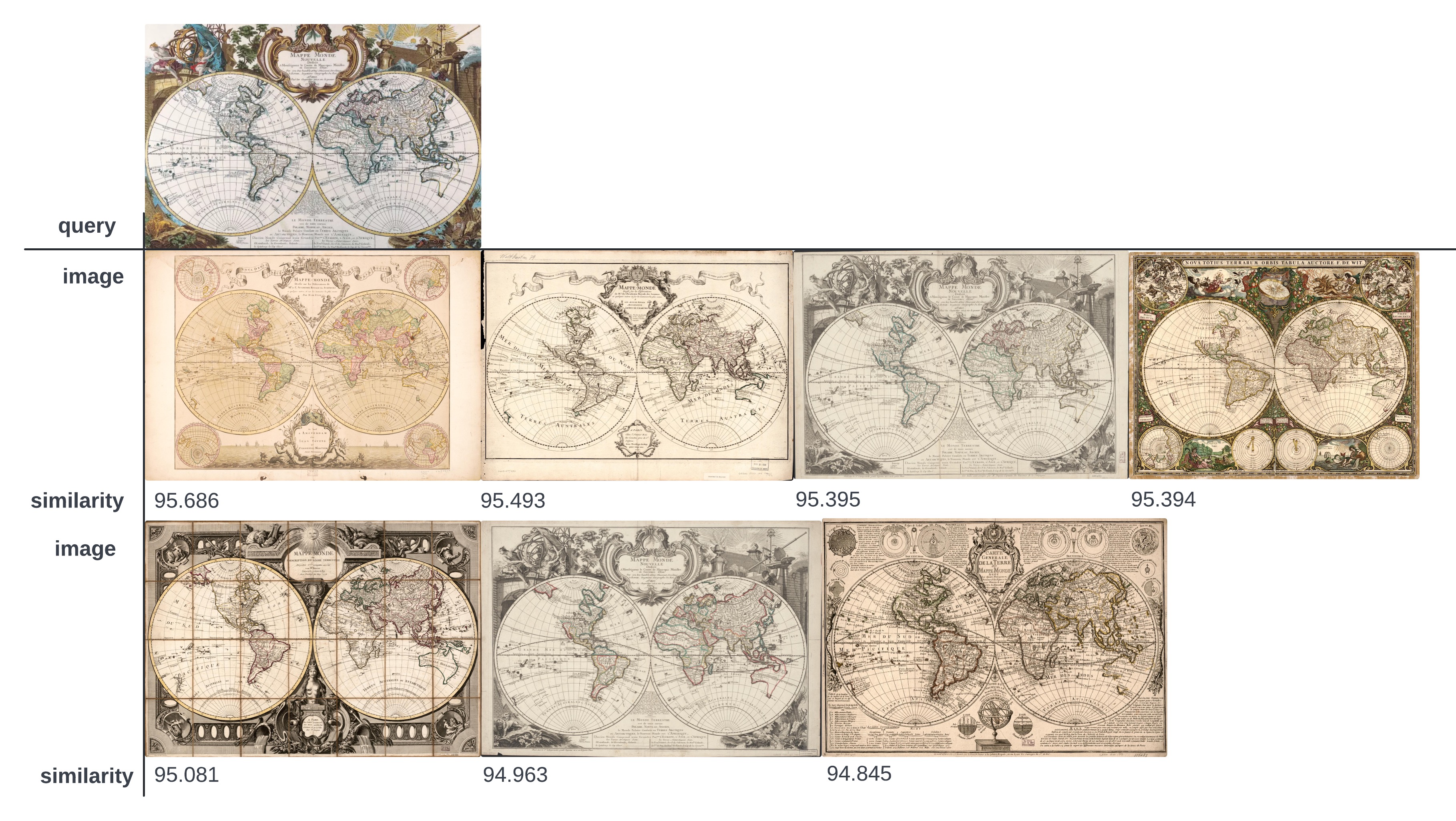} 
 \includegraphics[width = 0.74\linewidth ]{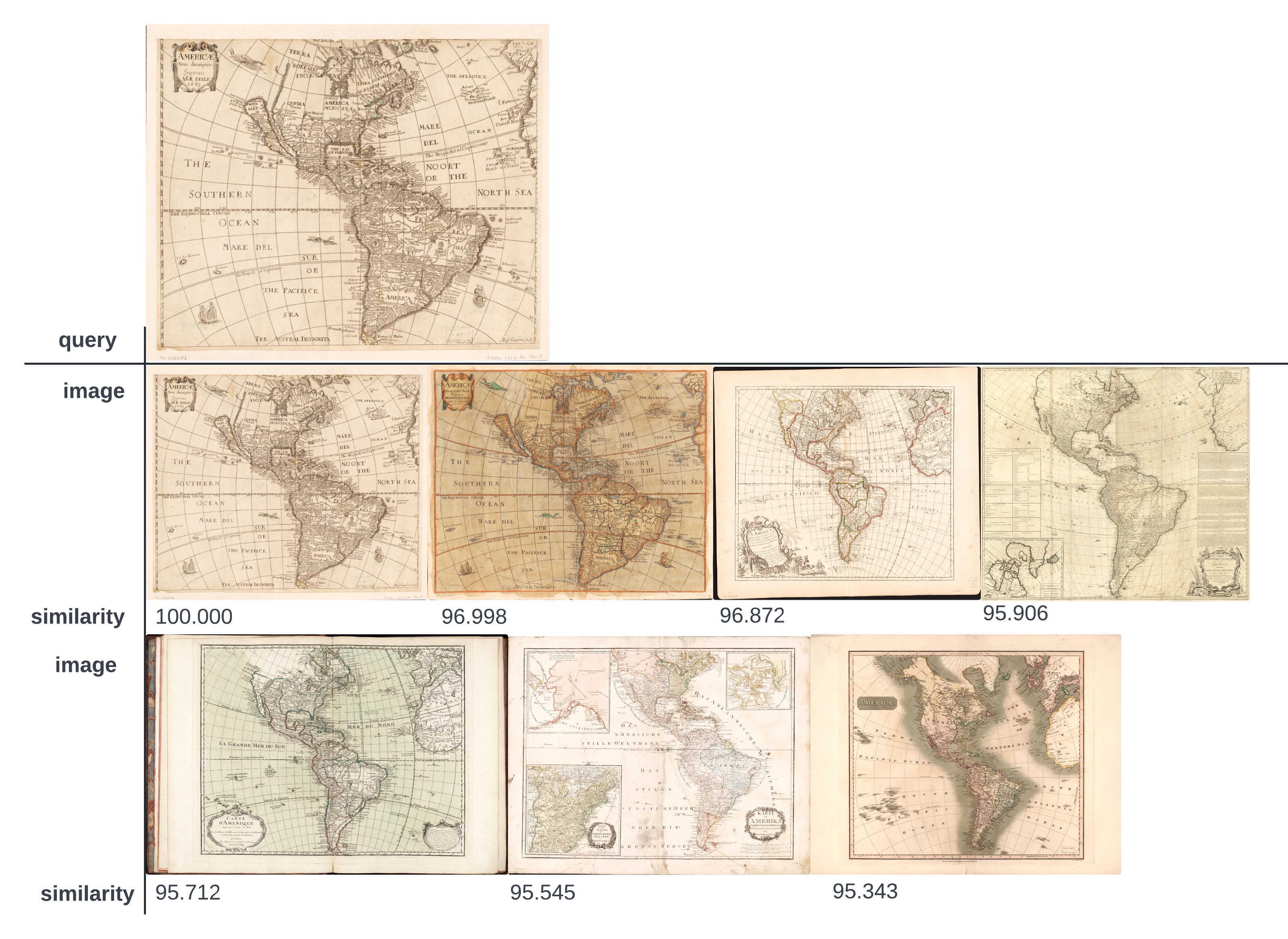} 
\end{center}
 \caption{The most relevant images returned by our search system for two different image inputs.}
\label{fig:results_image}
\end{figure}

\begin{figure}[pos=htbp]
\begin{center}
\textbf{Text- \& image-input Search}\\
 \includegraphics[width = 0.74\linewidth ]{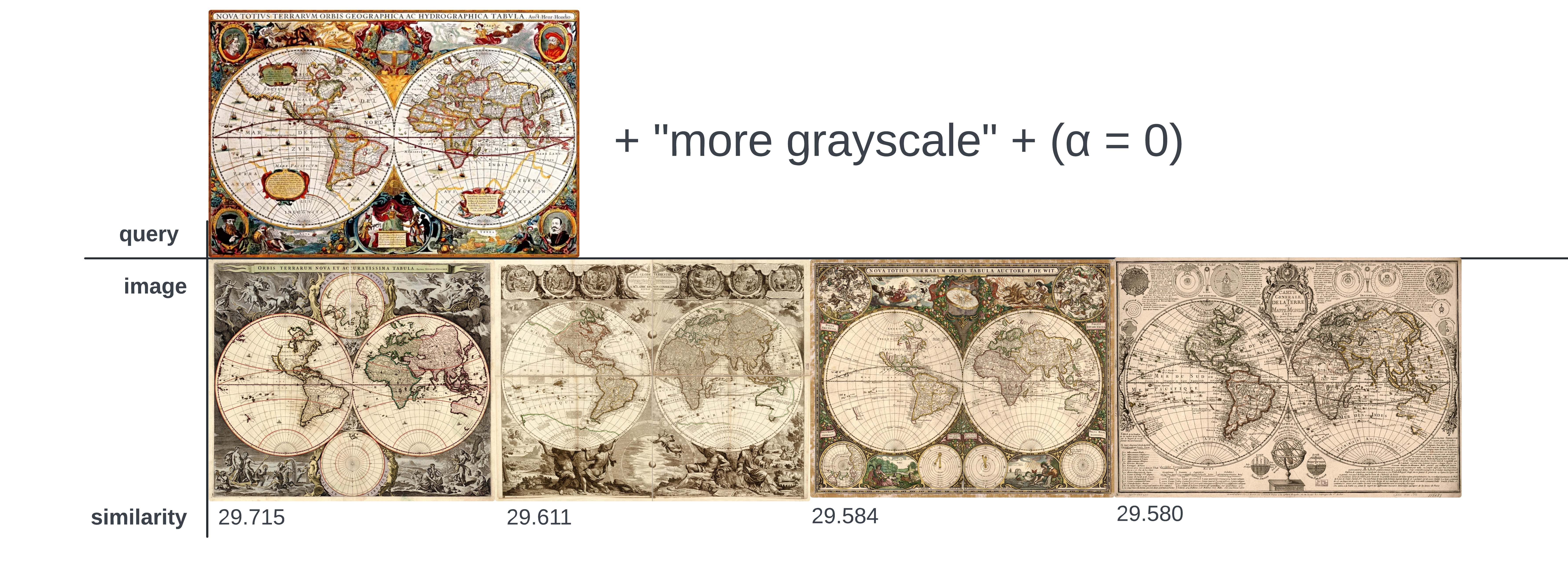} 
 \includegraphics[width = 0.74\linewidth ]{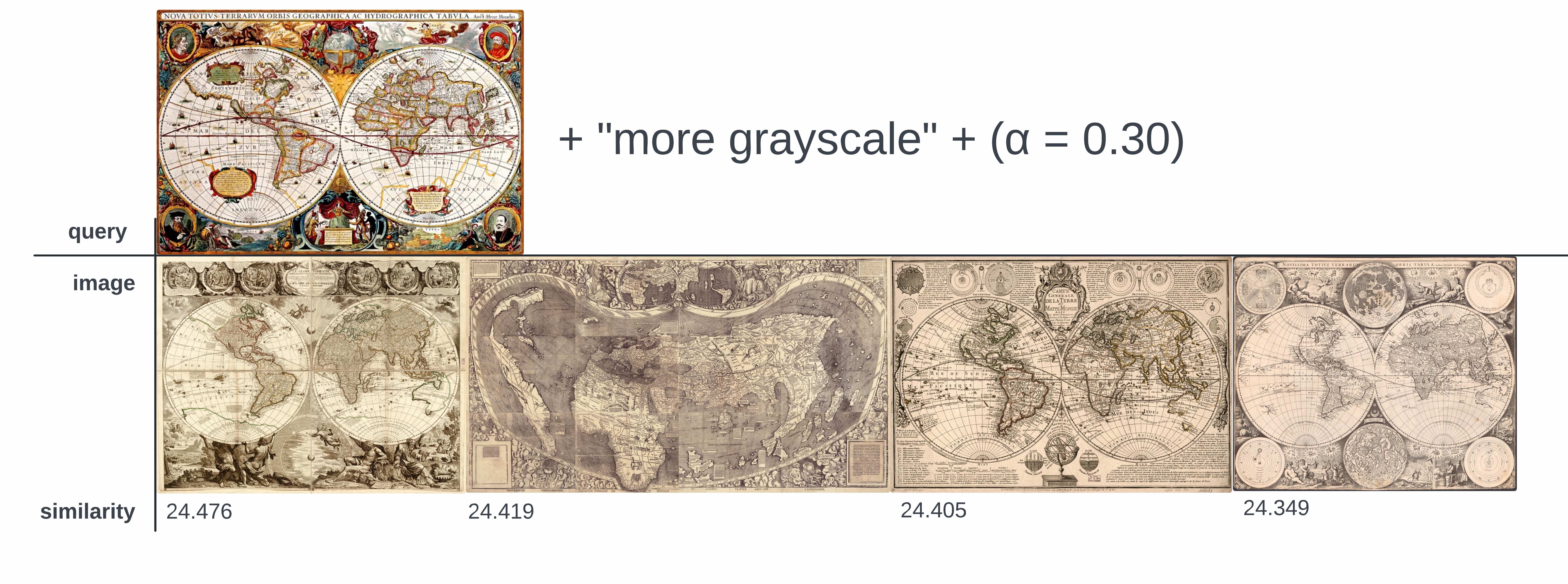} 
 \includegraphics[width = 0.74\linewidth ]{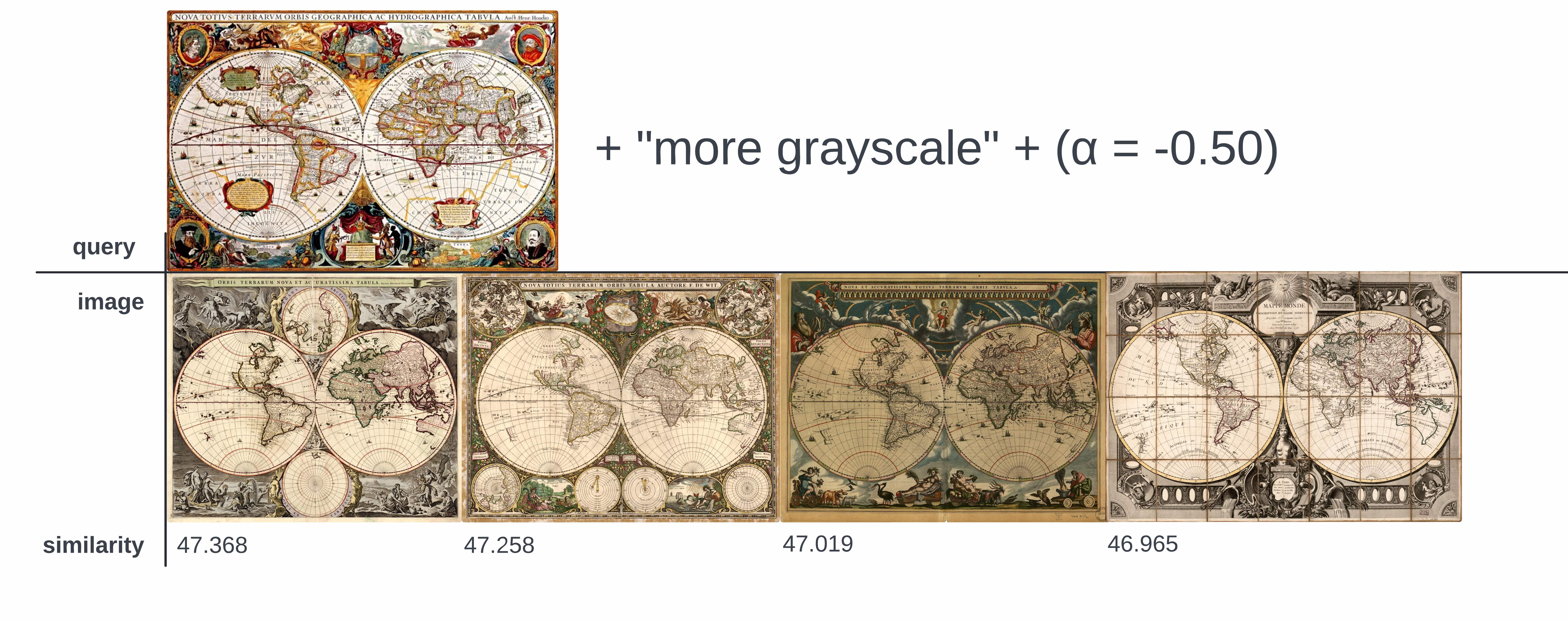} 
\end{center}
 \caption{The most relevant images returned by our search system when given the multimodal query of the specified image and the text guidance ``more grayscale'' with three different scaling factors ($\alpha = 0, 0.3,$ and $-0.5$, respectively). Note that as $\alpha$ is made more positive, the returned maps are more grayscale.}
 \label{fig:results_multimodal}
\end{figure}

\subsection{Search Results}

We begin by presenting example search results for all three approaches and reflecting on strengths and weaknesses. These observations are derived from our conversations with staff in the Library of Congress Geography and Map Division, who have experimented with our search implementation and have offered feedback based on their experiences with patron requests. As a brief observation in regard to the performance of our search implementation, we note that our search implementation processes a query, searches over all half a million images, and finds the most relevant results in less than a second on an M3 MacBook Pro with 18GB memory. This indicates that our implementation is both responsive and scalable.

\subsubsection{Text-input Search}\label{sec:results_description_text}

In Figure \ref{fig:results_text}, we show three examples of text-input searches with natural language queries: ``tattered and worn map,'' ``old panoramic map surrounded by images of buildings,'' and ``a map with illustrations of 19th-century ships.'' We chose these three examples in order to demonstrate a range of different searches that can be performed, from specific content within the maps (e.g., ships), to map styles (e.g., panoramic maps) and layouts (e.g., surrounded by images of buildings), to time periods (e.g., 19th-century), to the material properties of the maps (e.g., tattered and worn). 

Significantly, these approaches complement the metadata found in the catalog records for these maps. For example, a text search for ``celestial map'' yields eight relevant celestial charts out of the top ten results. Given that some but not all of the loc.gov JSON records for these maps include the word ``celestial'' (e.g., ``celestial chart,'' ``celestial sphere''), our search implementation enables the user to retrieve more relevant examples than what is possible when restricted solely to the existing metadata.

Conversely, a text search for ``map with cartouche'' yields more mixed results. Of the returned images, the second and third results are maps with cartouches, and the fourth result has a more general cartouche; however, other results do not have cartouches. A text search in \url{https://www.loc.gov/maps} for ``cartouche'' returned 199 results, which generally have cartouches depicted. In this case, existing metadata proves more useful (though it should be noted that our search implementation can easily be extended to include metadata search as well). Another example where metadata proves more useful is ``watercolor map.'' Examples where both our search implementation and metadata-based search do not have high precision include ``maps with drawings of people'' and ``hand drawn maps'' (though the latter could be partially constrained by searching terms in the metadata such as ``pencil,'' ``ink,'' or ``watercolor'').

\subsubsection{Image-input Search}\label{sec:results_description_image}

In Figure \ref{fig:results_image}, we show two examples of image-input, reverse-map searches. As with the text-input search examples, we have chosen these two maps to reflect distinct styles. This type of search can be useful in a number of settings. For example, a user may not know the proper vocabulary for specific visual features or styles, or relevant information may not be present in a map's metadata. As representative of a valuable use case, staff offered successful reverse-image searches with portolan charts -- a type of early nautical map that is visually recognizable by diagonal lines often referred to as rhumb lines or windrose lines. 

Additionally, though the text-input search of ``map with cartouche'' in Section \ref{sec:results_description_text} yielded mixed results, a reverse image search of a map with a cartouche performed significantly better (see the top search in Figure \ref{fig:results_image}), returning nine relevant maps out of the top ten maps returned for one example and seven relevant maps out of the top ten maps for another. In general, image-input searches typically yield more accurate results than text-based searches. Indeed, when measuring the similarity between query and results using raw cosine similarity scores (without applying softmax normalization), image-based searches achieve scores that are almost triple those of text-based searches.

A weakness with image-input search is that the user cannot constrain what in particular about the specified map input is most important to them. Indeed, this was a significant motivating factor for our implementation of joint text- \& image-input search.

\subsubsection{Text- \& Image-input Search}\label{sec:results_description_both}

In Figure \ref{fig:results_multimodal}, we introduce an example of joint text- \& image-input search with a map and the natural language query ``more grayscale,'' along with three different scaling factors: $\alpha = 0, 0.3,$ and $-0.5$. As with image-input search, this method could be particularly valuable when a user does not know the proper vocabulary, but this method offers the added affordance of enabling the user to specify natural language in order to tune the search. Because the user can quickly refine searches in an interactive fashion, we believe this affordance for specifying feedback is a promising one for exploratory search with maps.

Interestingly, a combined search of a map with a cartouche along with the text input ``map with cartouche'' yielded  better results than the equivalent searches with text-input only and image-input only. Using the two different example maps tested in Section \ref{sec:results_description_image}, along with the text input ``map with cartouche'' and a scaling factor of $\alpha = 0$, the inputs returned ten and nine relevant maps out of the top ten returned results, respectively. However, this method of reinforcing the search via both text and image did not improve results for other searches such as ``hand drawn maps,'' ``tattered map,'' or ``watercolor map,'' likely owing to the lack of additional information provided to the model across the two modes. 

\subsection{Applying the LC Labs AI Planning Framework}

Throughout our research, we have adhered to the LC Labs AI Planning Framework in order to engage with the responsible and ethical dimensions of this work \cite{planningframework}. We selected this framework in accordance with our use of Library of Congress materials. Created by LC Labs at the Library of Congress in 2023, the AI Planning Framework articulates three distinct elements (data, models, and people) across three phases (understanding, experimenting, and implementing) of a project's development \cite{labs_fw}. To offer relevant considerations and facilitate documentation during a project's development, the LC Labs AI Planning Framework provides three worksheets on data privacy and transparency \cite{planningframework}: 
\begin{enumerate}
    \item Use Case Risk Worksheet, ``to assist staff in assessing the risk profile of an AI use case.''
    \item Phase II Risk Analysis, ``to articulate success criteria, measures, risks, and benefits for an AI Use Case.''
    \item Data Readiness Assessment, ``to assess readiness and availability of data for the proposed use case.''
\end{enumerate}
We have completed all three worksheets and included them in our \href{https://github.com/j-mahowald/clip-loc-maps}{GitHub repository}. Based on our reflections during our completion of the worksheets, we note a few salient points. Because all training and search data are obtained from the Library of Congress, the overwhelming majority of the maps included are in the public domain (for any questions pertaining to a particular map's copyright, its included metadata can be consulted). We note that our fine-tuning dataset, taken directly from the larger corpus of maps described in Section \ref{sec:dataset}, is used to fine-tune the model and evaluate performance, as described in Section \ref{sec:finetuningdataset} and Section \ref{sec:finetuning} in the Appendix. In our worksheets, we describe the requirements and evaluations by Geography and Map staff at the Library of Congress, who serve as proxy evaluators for the intended end-use researchers. Lastly, we note that the absence of personally identifiable information, the low cost of mistakes in search and discovery, and the rigorous evaluations of our process make our application a low-risk use case according to the AI Planning Framework.

\section{Conclusion \& Future Work}

\subsection{Conclusion}

In this paper, we have asked a central question within the computational humanities: how might emerging methods from multimodal machine learning be utilized to facilitate  searching large-scale map collections? To address this question, we have built out a search implementation for 562,842 images of maps publicly available via the Library of Congress's API. In particular, we have produced CLIP embeddings for all 562,842 images and introduced a search implementation that enables three different kinds of search inputs: 1) natural language, text-based inputs, 2) visual, image-based inputs, and 3) multimodal, combined text- and image-based inputs. In consultation with staff in the Library of Congress Geography and Map Division, we have explored example searches and demonstrated the utility of these search methods. Moreover, we have demonstrated a commitment to responsible and ethical AI practices by following the LC Labs AI Planning Framework. For further work with historic maps, we have released a dataset of 10,504 map-caption pairs, along with an architecture for fine-tuning CLIP on this dataset. To facilitate transparency and re-usability for our code by end-users such as scholars and practitioners, we have released all of our code into the public domain as Jupyter notebooks. In what remains, we explore the extensibility of our approaches to other digital collections held by galleries, libraries, archives, and museums, as well as describe other future work.

\subsection{Toward Improved Discoverability in GLAM Collections with CLIP}\label{sec:toward}

Digital collections continue to grow at enormous rates. Developing methods of facilitating search and discovery are more important than ever in order to contend with the challenge of scale. Our search implementation leveraging CLIP has demonstrated the potential for searching maps beyond their catalog records and existing metadata. Here, natural language inputs facilitate interactive navigation, an important component of exploratory search. 

With the marked improvements in multimodal machine learning over the past few years, it is clear that there are manifold opportunities to improve access through the application of these methods. Significantly, these methods are extensible to a wide range of digitized and born-digital collections currently only searchable via metadata and text search. In the case of born-digital collections such as web archives, the lack of structured metadata at the webpage level necessitates the exploration of these methodologies. Example searches that would be enabled by applying CLIP-like approaches are as far-ranging as finding heavily redacted pages in born-digital government documents to identifying specific motifs in rare book illustrations. We have shown that our implementation can render half a million images searchable on a single laptop, demonstrating that such approaches are scalable to millions of items with little modification.

The application of these methodologies presents challenges as well. Digital collections are incredibly heterogeneous, spread across time periods, languages, media types, and beyond, with different metadata fields. Consequently, ensuring that these approaches surface relevant facets, and doing so responsibly and ethically, must be primary considerations for this work. As one example, developing fine-tuned models for this work is important, but recognizing the limitations and failure modes of these approaches -- and when to use machine learning-based approaches to begin with -- is just as important. We therefore advocate that researchers continue to adopt frameworks such as the LC Labs AI Planning Framework during all stages of a project when applying AI to digital collections.

\subsection{Future Work}\label{sec:future-work}

Many directions of future work remain of interest for us. First, we would like to continue the fine-tuning experiments described in Section \ref{sec:finetuningdataset}, as well as Section \ref{sec:finetuning} of the Appendix. Though our initial experiments showed mixed results, we believe additional experiments surrounding careful implementation of a fine-tuned model to the search engine could reduce some of the noise from inaccurate searches. Indeed, given the demonstrated utility of fine-tuned embeddings for a range of downstream tasks in machine learning more generally, we believe that a fine-tuned CLIP model for historic maps in particular could be beneficial to the computational humanities community. Along these lines, we are interested in exploring additional approaches to training and fine-tuning multimodal models, such as ones that do not utilize contrastive learning and are not restricted by the contrastive fine-tuning mechanism \cite{multimodal_survey}.

Moreover, we plan to build a proper search interface for our implementation with the goal of hosting an exploratory search system that can be publicly accessed. Given the importance of front-end affordances and considerations from human-computer interaction, we believe detailed analysis surrounding the best interaction mechanisms warrants further study \cite{Hearst_2009}. This is especially important for maps, where affordances for browsing must take into consideration the specificity of viewing and interacting with the digital objects themselves, which are often large and often span multiple sheets \cite{mapscholar}. User studies would be beneficial for building a system that would be most valuable to patrons. We also plan to incorporate metadata search into this interface, with the understanding that combining our search implementation with existing search fields would be complementary.

Lastly, as described in Section \ref{sec:toward}, we believe that these multimodal, CLIP-style approaches to search and discovery are useful for a wide range of digital collections. As a result, we have begun exploring extensibility to other document types including web archives, born-digital documents, digitized books, and digitized newspapers. Indeed, given the ongoing, worldwide efforts surrounding the creation and stewardship of both digitized and born-digital collections, continuing to refine methods for improving discoverability will only grow in importance over the coming years.

\begin{acknowledgments}
This work was supported in part by the Library of Congress Kluge Fellowship in Digital Studies (BCGL), as well as the Archer Fellowship Program and the University of Texas System (JM). At the Library of Congress, we would like to thank Rachel Trent, Amelia Raines, and Sundeep Mahendra in the Geography and Map Division for their many-month collaboration surrounding querying the digitized maps held by the LC, as well as their invaluable advice and input surrounding relevant searches for map patrons. We would also like to thank Abigail Potter and Brian Foo in LC Labs for their guidance surrounding the Library of Congress's AI Planning Framework. Lastly, we thank Michael Stratmoen and Travis Hensley in the Kluge Center for their support of the internship and fellowship programs that made this collaboration and paper possible.
\end{acknowledgments}

\bibliography{bibliography}

\appendix
\section{Appendix: A Description of Experiments for Fine-tuning CLIP}\label{sec:finetuning}

This section serves to describe our experiments fine-tuning CLIP on map-caption pairs, which have yielded mixed results to date. In our fine-tuning script available in our \href{https://github.com/j-mahowald/clip-loc-maps/tree/main/fine-tuning}{repository}, we initialize the pre-trained model and processor as introduced in the Hugging Face transformers library (\texttt{openai/clip-vit-base-patch32}) \cite{hf_clip}. We define an image-text pair dataset inheriting from the PyTorch Dataset class, into which we load lists of image paths and their corresponding captions. This dataset is then fed into a PyTorch DataLoader with a custom collate function that opens images, converts them to RGB, and uses the CLIP processor to batch and preprocess texts and images together. We then define the optimizer under Adam with traditional hyperparameters (\(\beta_1 = 0.9, \beta_2 = 0.98, \epsilon = 1 \cdot 10^{-6}, \lambda = 0.2\)) in Kingma \& Ba \cite{adam} and a scheduled learning rate.

The script initiates a 16-epoch training loop. Each epoch processes \(15000 // 32 + 1 = 469 \) batches of 32 image-text pairs, where gradient computations are reset and inputs are prepared and passed through the model for each batch. The model computes logits for images and texts and calculates a symmetric InfoNCE (noise-cross estimation) loss, a common choice for contrastive learning like with CLIP, which encourages the model to align the embeddings of matching texts and images while distinguishing non-matching ones. InfoNCE loss is selected against cross-entropy loss, on which the original model is trained, in light of the increased ``noise'' associated with the larger caption pool in our fine-tuning set. The gradients are backpropagated, and the optimizer updates the model's weights. After each epoch, the average and total losses are calculated and recorded. The model state, optimizer state, and training loss are saved to a checkpoint file, allowing for training or evaluation to be resumed later.

For our first training regimen, we selected a random sample of \(n=50,000\) standalone image-text pairs to generate and feed into the fine-tuning model. Because \texttt{loc.gov} metadata are typically written at the item level, we limited our sample to images that are part of items with fewer than 10 segments to avoid vague or potentially inaccurate metadata-derived captions. This configuration resulted in a loss reduction of about 50\% over 16 epochs, with the logarithmic decline suggesting that the average error would decrease below one only after 35 to 40 epochs. We choose neither to pursue this path nor to increase the dataset size out of caution for overfitting (a training set of 50,000 already represents 10\% of the entire collection) and for potential inaccuracies in the metadata-derived caption generation. The accuracy for the specific task of search and discovery across the entire collection suffers qualitatively with this fine-tuned model as compared to the base model. 

For our second training regimen, we utilize the dataset introduced in Section \ref{sec:dataset}. From the first iteration, we recognized that the broader map collection is heavily skewed toward the Division's collection of Sanborn fire insurance maps, which biased the fine-tuning data. The dataset of $10,504$ map-captions was constructed with this consideration in mind. We then record the loss reduction across five epochs for an array of learning rates and batch sizes, finding little significant reduction (and, at times, increase) in validation loss. This occurred across several (8, 16, 32, 64) batch sizes and learning rates, though the decline was more modest for smaller batch sizes. This phenomenon owes partially to the regressive nature of machine learning for search and discovery. Whereas contrastive models tasked with classifying across a discrete set of outputs (for instance, a list of possible years during which a photograph was taken \cite{wevers_photos}) largely benefit from several cycles of supervised learning, the infinite label space of a regression problem requires the model to interpolate or extrapolate beyond the finite set of examples seen during training.

\end{document}